\documentclass[onecolumn,amsmath,amssymb,nofootinbib,aapt]{revtex4} 

\usepackage{amsmath}
\usepackage{graphicx}
\usepackage{graphics}
\usepackage{dcolumn}
\usepackage{bm}
\usepackage{hyperref}
\usepackage{xcolor}

\def\BEq{\begin{equation}}
\def\EEq{\end{equation}}
\def\BEqA{\begin{eqnarray}}
\def\EEqA{\end{eqnarray}}
\def\BEn{\begin{enumerate}}
\def\EEn{\end{enumerate}}
\def\BWT{\begin{widetext}}
\def\EWT{\end{widetext}}

\begin{document}

\title{Exotic spherically-symmetric $\Lambda$-vacuum in 
the four-dimensional Starobinsky model}

\author{Andrei Galiautdinov}
\email{ag1@uga.edu}
\affiliation{
Department of Physics and Astronomy, 
University of Georgia, Athens, Georgia 30602, USA
}

\date{\today}

\begin{abstract}
We introduce an exact, two-parameter family of static, 
spherically-symmetric, constant-curvature $\Lambda$-vacuum 
solutions within the four-dimensional Starobinsky 
$f(R)=R+\alpha R^2+2\Lambda$ model. When the bare cosmological 
constant is precisely fine-tuned to $\Lambda = 1/(8\alpha)$, 
the scalar curvature is rigidly fixed such that the derivative 
$f'(R)=1+2\alpha R$ identically vanishes. Because this derivative 
acts as the effective multiplier for the standard curvature terms
in the modified field equations, its global vanishing mathematically 
erases the normal rules of gravitational dynamics, demonstrating 
that the family represents a pathological boundary to the space 
of viable physical geometries. This exact decoupling of the field equations 
permits the existence of a fundamentally unconstrained $1/r^2$ integration 
constant in the metric, which functions as a purely geometric 
Reissner-Nordstr\"om hair mimicker. However, any infinitesimal classical 
deviation from this exact boundary instantaneously destroys the degeneracy, 
rigorously forcing the geometric hair to vanish and discontinuously collapsing 
the spacetime back into the standard, dynamical Schwarzschild-de Sitter solution. 
We provide the exact derivation of this spacetime and methodically 
highlight its physical pathologies, including the identically vanishing 
Wald entropy of the associated black hole horizons, the strict divergence 
of the effective gravitational coupling, the complete breakdown of the 
test-particle approximation, and the onset of severe ghost instabilities. 
Ultimately, this exact solution 
functions as a ``do not enter'' sign within the Starobinsky 
model, pedagogically illustrating the extreme fragility and 
physical hostility of degenerate, purely mathematical solutions 
in highly non-linear $f(R)$ gravity theories.
\end{abstract}

\maketitle


\section{Introduction}

The observational imperatives of modern cosmology, 
particularly the phenomena of early-universe inflation and 
late-time accelerated expansion, have driven extensive 
theoretical interest in infrared modifications of general 
relativity. Among the most natural and thoroughly investigated 
extensions are $f(R)$ gravity theories, which introduce 
higher-order curvature invariants into the Einstein-Hilbert action 
(e.g., \cite{Capozziello2008, Capozziello2011, Nojiri2007, Nojiri2011, 
Nojiri2017, DeFelice2010, Sotiriou2010, Alvarez-Gaume2015}). 
The canonical archetype of such theories is the Starobinsky 
model, defined by the Lagrangian density $f(R) = R + \alpha R^2$, 
which was initially proposed to generate a non-singular de Sitter 
phase in the early universe and naturally drive cosmic inflation 
\cite{Starobinsky1980, Starobinsky1983, Vilenkin1985}. 

While the cosmological successes of the Starobinsky model are well 
known, testing the viability of any modified gravity theory necessitates 
a rigorous understanding of its strong-field regime, specifically in the 
context of compact astrophysical objects and black holes. Consequently, 
the search for static, spherically-symmetric solutions in $f(R)$ gravity 
has become a major theoretical endeavor 
\cite{Multamaki2006, Nelson2010, Sebastiani2011, 
Lu2015, Yu2018, Cikintoglu2018, Podolsky2020} 
(also, cf.\ \cite{Frolov2009}). 

However, because the field equations of fourth-order gravity are 
highly non-linear and coupled, finding exact analytical solutions 
beyond the trivial Schwarzschild or Schwarzschild-de Sitter geometries 
has proven notoriously difficult.
Early systematic investigations, such as the comprehensive analysis 
in Ref.\ \cite{Carames2009}, 
have demonstrated that the existence of exact analytical 
solutions is strongly dependent on the specific algebraic 
structure and constraints placed upon the derivative function 
$F(R) = f'(R)$. In general $f(R)$ models, including 
higher-dimensional extensions, it is precisely the behavior 
of this derivative function that governs the dynamic degrees 
of freedom. Foundational investigations into static, 
spherically-symmetric black holes in modified gravity, such 
as \cite{Dombriz2009}, have established the rigorous conditions 
under which $f(R)$ theories admit constant-curvature solutions. 
These studies demonstrated that constant scalar curvature spacetimes 
generically reduce to the standard Schwarzschild-de Sitter geometry, 
provided the background curvature satisfies the vacuum trace constraint. 
This standard uniqueness characterizes the physical, dynamical 
branch of these theories, wherein the derivative of the Lagrangian 
density remains strictly non-zero ($f'(R) \neq 0$). 
Building upon this solid foundation, the present work focuses on 
the singular complement to this established regime:   
the exact, non-dynamical mathematical boundary where 
$f'(R) = 0$. 

As discussed by Kehagias et al.\ \cite{Kehagias2015}, 
in pure $R^2$ gravity where $f(R) = R^2$, any spacetime with 
a vanishing scalar curvature ($R=0$) yields $f'(R)=0$, which 
functionally decouples the trace of the field equations. 
This loss of dynamics leads to a massive mathematical degeneracy 
that permits exact solutions with arbitrary $1/r^2$ integration 
constants. Rather than representing \emph{physical} geometric 
``hair,'' such terms are artifacts of the decoupled equations, manifesting 
exactly where the standard predictability established in 
\cite{Dombriz2009} breaks down.

When supplementing the Starobinsky model with a bare 
cosmological constant, $f(R)=R + \alpha R^2 + 2\Lambda$, 
the presence of the linear Einstein-Hilbert term generally 
avoids this $R=0$ degeneracy, as $f'(0) = 1 \neq 0$. 
The primary pedagogical motivation of this note is to 
demonstrate that an analogous mathematical decoupling 
can be systematically recovered at a finite background curvature 
in the four-dimensional Starobinsky model by the inclusion of 
a finely tuned bare cosmological constant, $\Lambda$. 
In this singular regime, there arises a two-parameter 
family of static, spherically-symmetric, constant-curvature 
$\Lambda$-vacuum solutions. We show that when the cosmological 
constant is fixed to $\Lambda = 1/(8\alpha)$, the scalar curvature 
of the spacetime is rigidly forced to $R = -1/(2\alpha)$, which is precisely 
the root of the derivative $f'(R)=1 + 2\alpha R = 0$. 
By carefully navigating the modified field equations at this locus, we show 
that this constraint enables the exact integration of a mathematically degenerate 
spacetime geometry. This solution features an unconstrained $1/r^2$ metric term, acting 
as a purely geometric mimicker of physical charge, alongside a formal mass parameter 
(which would be identified with the standard Newtonian mass only in the well-behaved, 
non-degenerate regime).

We emphasize that the emergence of such fine-tuning is 
not an isolated anomaly, but rather a generic algebraic 
inevitability of constant-curvature $f(R)$ gravity. In any 
generic $f(R)$ theory, imposing the condition that the 
scalar curvature is strictly constant ($R = A$) forces all 
derivative terms in the vacuum field equations, 
\BEq
f'(R)R_{\mu \nu }
-{\frac {1}{2}}g_{\mu \nu }f(R)
-\left[\nabla _{\mu }\nabla _{\nu }- g_{\mu \nu }\Box \right]f'(R)
=0,
\EEq
to identically vanish. These general fourth-order equations 
automatically collapse to the second-order form,
\begin{equation}
f'(A) R_{\mu\nu} - \frac{1}{2} f(A) g_{\mu\nu} = 0.
\end{equation}
Taking the trace yields the general master constraint for 
constant-curvature backgrounds,
\begin{equation}
A f'(A) - 2f(A) = 0.
\end{equation}
By isolating $f(A)$ from this trace constraint and substituting 
it back into the reduced field equations, one finds that any 
constant-curvature metric in $f(R)$ gravity must satisfy the 
factorized equation,
\begin{equation}
f'(A) \left( R_{\mu\nu} - \frac{A}{4} g_{\mu\nu} \right) = 0.
\end{equation}
For the Starobinsky action, $f(R) = R + \alpha R^2 + 2\Lambda$, 
the strictly quadratic nature of the theory renders the master 
trace constraint perfectly linear in $A$. This rigidly fixes the 
background curvature to $A = -4\Lambda$ and yields the 
fine-tuning prefactor $f'(A) = 1 - 8\alpha\Lambda$, as 
derived in our main Eq.\ (\ref{eq:4dStaroMODconstantR}) 
below. Constructing a solution at the degenerate $f'(A)=0$ 
boundary is then a matter of judiciously selecting 
a constant-curvature Ansatz that best exposes the physical 
consequences of this mathematical decoupling, rather than 
arbitrarily searching for novel or ``exotic'' geometries.
 
The prospect of Reissner-Nordstr\"om-like geometries emerging 
in extended gravity has long been a subject of considerable interest. 
For instance, foundational work by Cembranos et al.\ \cite{Cembranos2011} 
established that exact Reissner-Nordstr\"om and Kerr-Newman 
black holes exist as valid, dynamical solutions in generic $f(R)$ theories, 
provided the spacetime is explicitly coupled to a standard Maxwell 
tensor and the constant scalar curvature satisfies the modified trace 
equation. In stark contrast, the exact geometry we present here is 
fundamentally distinct: our spacetime is strictly a $\Lambda$-vacuum, 
entirely devoid of any electromagnetic source. The arbitrary $1/r^2$ 
term in this degenerate regime is a purely mathematical artifact, emerging 
exclusively because the finely-tuned condition $f'(R)=0$ 
decouples the field equations. Thus, this solution serves as an exact, 
analytical cautionary tale that demonstrates how higher-order curvature 
self-interactions, when pushed to a singular limit, can spontaneously 
generate a metric that perfectly mimics a physically charged black 
hole without the presence of an actual $U(1)$ gauge field.

To cement this pedagogical warning, we systematically outline 
the physical and thermodynamic pathologies of this mathematically 
exact background. Because the spacetime exists exclusively at the $f'(R)=0$ 
boundary, the effective gravitational coupling strictly diverges, 
leading to a complete breakdown of the test-particle approximation.
Any physical probe with non-zero mass would theoretically induce 
an infinite backreaction, rendering the spacetime entirely 
fragile against matter perturbations. Alongside this backreaction
catastrophe, we draw upon established formalisms for black hole 
thermodynamics in modified gravity \cite{Briscese2008, Volovik2024} 
to show that horizons in this spacetime possess identically 
vanishing Wald entropy. Rather than a mere paradox, 
this result acts as a vital thermodynamic consistency check, perfectly 
reflecting the macroscopic loss of dynamical degrees of freedom and 
underscoring the inherently pathological nature of this fine-tuned 
Starobinsky vacuum.

\section{Derivation}

We begin by considering the action for the four-dimensional 
Starobinsky model supplemented by a bare cosmological 
constant, $\Lambda$. In the Jordan frame, the action is given 
by
\begin{equation}
S = -\frac{1}{16\pi G}\int d^4 x \sqrt{-g} 
\left( R + \alpha R^2 + 2\Lambda \right),
\end{equation}
where $G$ is the bare Newton's constant and $\alpha$ is 
the Starobinsky parameter governing the strength of the 
quadratic curvature correction. We use the metric signature 
$(+---)$ and employ the sign convention for the cosmological 
constant adopted by Zeldovich (see Appendix I, Eq.\ (I.1), 
Ref.\ \cite{Zeldovich1968}). Varying\footnote{Here we operate 
strictly within the metric formalism, wherein the action is varied 
exclusively with respect to the metric tensor $g_{\mu\nu}$, 
treating the connection as the standard Levi-Civita connection. 
This choice uniquely yields the fourth-order field equations 
characteristic of metric $f(R)$ gravity. 
Strictly speaking, since the exact solution 
presented below resides on the mathematically degenerate 
boundary where $f'(R)=0$, the standard variational principle 
becomes ill-posed because the generalized Gibbons-Hawking-York 
boundary term (e.g., \cite{Dyer2009}), which is proportional 
to $f'(R)$, identically vanishes. Our approach effectively 
relies on deriving the field equations in the generic dynamical domain 
where $f'(R) \neq 0$, and then formally working with the exact 
limit of these continuous equations at the boundary. 
This underscores the central physical thesis of the work: the resulting 
geometry represents a highly pathological state devoid of standard 
dynamical degrees of freedom.} this action with 
respect to the metric $g_{\mu\nu}$ yields the fourth-order 
$\Lambda$-vacuum field equations,
\begin{equation}
\label{eq:4dStaro}
\left( R_{\mu\nu} -\frac{1}{2}Rg_{\mu\nu} -\Lambda g_{\mu\nu}\right)
+2\alpha R \left( R_{\mu\nu} -\frac{1}{4}Rg_{\mu\nu}\right)
-2\alpha \left( R_{;\nu\mu} - g_{\mu\nu} R^{;\lambda}_{\; ;\lambda}\right)
=0.
\end{equation}
The dynamics of the scalar curvature can be isolated by taking 
the trace of Eq.\ (\ref{eq:4dStaro}), which provides the scalar 
constraint,
\begin{equation}
\label{eq:scalarConstraint}
2\alpha R^{;\lambda}_{\; ;\lambda}=\frac{R+4\Lambda}{3}.
\end{equation}
By substituting this trace relation back into the original field 
equations to eliminate the $R^{;\lambda}_{\; ;\lambda}$ term, 
the system can be rewritten in a more mathematically tractable 
form,
\begin{equation}
\label{eq:4dStaroMODv2}
(1+ 2\alpha R) R_{\mu\nu} 
+\frac{2\Lambda -(1+ 3\alpha R) R}{6} g_{\mu\nu}
-2\alpha R_{;\nu\mu} 
=0.
\end{equation}

To search for black hole solutions, we adopt the simplest static, 
spherically-symmetric Ansatz,
\begin{equation}
\label{eq:hAnsatz}
ds^2=h(r)dt^2
-\frac{dr^2}{h(r)}
- r^2 d\Omega^2,
\end{equation}
for which the scalar curvature has the form,
\begin{equation}
R=h''(r)+\frac{2 \left(2 r h'(r)+h(r)-1\right)}{r^2}.
\end{equation}
If we demand that the curvature is constant, $R=A$, this 
differential equation can be directly integrated to yield 
the general local solution,
\begin{equation}
\label{eq:h(r)}
h(r)= 1 +\frac{c_1}{r}+\frac{c_2}{r^2}+ \frac{A}{12} r^2,
\end{equation}
where $c_1$ and $c_2$ are arbitrary constants of integration,
and
\BEq
\label{eq:A}
A=-4\Lambda,
\EEq
on the basis of (\ref{eq:scalarConstraint}). Correspondingly, the
field equations (\ref{eq:4dStaroMODv2}) simplify substantially, 
becoming
\begin{equation}
\label{eq:4dStaroMODconstantR}
(1-8\alpha \Lambda) 
\left( 
R_{\mu\nu} +\Lambda  g_{\mu\nu}
\right)
=0,
\end{equation}
which, as we pointed out in the Introduction, was expected
on general grounds. This generates two strictly distinct families 
of static solutions:

\begin{enumerate}
\item \textbf{The standard Schwarzschild-de Sitter family:} If we 
assume the dynamical regime $1-8\alpha \Lambda \neq 0$, then we recover the 
standard general relativistic 
vacuum field equations with a cosmological constant,
\BEq
\label{eq:standardGRfieldEqs}
R_{\mu\nu} +\Lambda  g_{\mu\nu} = 0.
\EEq
Plugging (\ref{eq:h(r)}) subject to (\ref{eq:A}) into 
(\ref{eq:standardGRfieldEqs}) results 
in a single algebraic constraint,
\begin{equation}
\label{eq:c2constraint}
\frac{c_2}{r^4}  = 0,
\end{equation}
rigidly forcing $c_2 = 0$. Identifying $c_1 = -2M$ as 
the physical mass parameter, we recover the familiar, well-behaved geometry,
\begin{equation}
h(r)= 1 -\frac{2M}{r} - \frac{\Lambda}{3} r^2,
\quad
\Lambda \neq \frac{1}{8 \alpha}.
\end{equation} 

\item \textbf{The mathematically degenerate boundary 
limit (the ``exotic'' family):} A critical 
mathematical singularity arises if $1-8\alpha \Lambda = 0$. 
In this exact limit, $1 + 2\alpha A = 0$ and, consequently, 
$A + \alpha A^2 + 2 \Lambda = 0$ as well. 
Physically, these expressions represent the derivative 
of the Lagrangian with respect to the scalar curvature, 
$f'(R) = 1 + 2\alpha R$, and the Lagrangian itself, 
$f(R)=R+\alpha R^2 + 2 \Lambda$. 
When the derivative vanishes, the dynamic tensor constraints on 
the Ricci tensor collapse entirely (Eq.\ \ref{eq:4dStaroMODconstantR} 
becomes identically $0=0$). 
Consequently, the only remaining 
constraint on the geometry is the scalar trace equation $R=A$, 
which is unconditionally satisfied by the general metric ansatz 
for any arbitrary value of $c_2$.
Denoting \emph{formally} $c_1 = -2M$, we 
obtain the exact degenerate solution,
\begin{equation}
h(r)= 1 -\frac{2M}{r}+\frac{c_2}{r^2}-\frac{r^2}{24\alpha}, 
\quad
\text{with} \quad
\Lambda = +\frac{1}{8 \alpha},
\end{equation} 
and $c_2$ fundamentally unconstrained. 
Thus, at this mathematically degenerate 
boundary where the bare, fine-tuned cosmological constant perfectly 
balances the quadratic coupling, the Starobinsky vacuum 
supports an arbitrary $1/r^2$ metric artifact mimicking 
physical geometric hair.
\end{enumerate}

We note an important algebraic triviality: if one sets $c_2 = 0$, 
the degenerate metric identically recovers the standard 
Schwarzschild-de Sitter geometry. 
However, a strict distinction must be drawn between 
geometric equivalence 
and dynamical equivalence. Even without the arbitrary 
$1/r^2$ term, this specific configuration resides exactly 
at the $\Lambda = 1/(8\alpha)$ locus, meaning the condition 
$f'(R) = 0$ persists globally. Consequently, this ``hairless'' subset of the 
degenerate branch remains fundamentally pathological, suffering 
from the identical divergent effective coupling and vanishing Noether 
charges as the $c_2 \neq 0$ spacetimes. It is dynamically distinct 
and physically disjoint from the healthy Schwarzschild-de Sitter 
family derived in the 
dynamical $1-8\alpha\Lambda \neq 0$ regime.

\section{Physical Pathologies and Thermodynamic Consistency}
The exact degenerate solution derived in the previous section 
represents a highly pathological mathematical configuration with instructive 
physical and thermodynamic consequences. 
In this section, we methodically detail the unphysical nature of the integration 
constant $c_2$, the identically vanishing black hole entropy, 
and the severe, catastrophic instability of the spacetime.

\subsection{The Integration constant $c_2$ and 
the Reissner-Nordstr\"om mimicker}

A notable artifact of the mathematically degenerate branch 
is the emergence of the $c_2 / r^2$ term in the metric function 
$h(r)$. In the standard framework of general relativity, a $1/r^2$ 
contribution to the metric potential arises exclusively from 
the stress-energy tensor of a $U(1)$ gauge field, yielding 
the classic Reissner-Nordstr\"om electro-vacuum solution 
where $c_2 \propto Q^2$, with $Q$ being the electric or 
magnetic charge of the black hole. 

However, in the fine-tuned Starobinsky $\Lambda$-vacuum 
presented here, there is no Maxwell field. The $c_2 / r^2$ 
term is generated entirely by the loss of the trace constraint 
in the higher-derivative field equations. Because $c_2$ is 
a fundamentally unconstrained integration constant of these 
decoupled differential equations, it represents a purely 
mathematical artifact rather than a \emph{physical} geometric ``hair.'' 
Hypothetically speaking, an external observer probing 
the formal metric of this spacetime 
would encounter a geometry resembling 
a Reissner-Nordstr\"om-de Sitter black hole, yet the spacetime 
is strictly devoid of electric charge. This provides an exact pedagogical 
example of how higher-order gravity, when pushed to a singular 
parameter locus, can produce metric terms that perfectly mimic standard 
matter fields. Furthermore, unlike the strictly positive $Q^2$ 
of the Maxwell field, the geometric parameter $c_2$ can 
formally take both positive and negative values, further highlighting 
its unphysical origin and allowing for formal geometries with no 
standard general relativistic counterpart.

\subsection{Reduction to pure quadratic gravity}

The existence of this degenerate exact solution can be understood 
more transparently by examining the structure of 
the Lagrangian at the $f'(R)=0$ boundary. Because the Starobinsky 
action $f(R) = R + \alpha R^2 + 2\Lambda$ is strictly a second-degree 
polynomial in the scalar curvature, its Taylor expansion around 
the background $R_0 = -1/(2\alpha)$ truncates exactly at the second order. 
The action can therefore be rewritten as
\begin{equation}
f(R) 
= 
f(R_0) + f'(R_0)(R-R_0) + \frac{1}{2}f''(R_0)(R-R_0)^2 
= 
\alpha(R-R_0)^2.
\end{equation}
At this specifically tuned boundary, the linear Einstein-Hilbert term and 
the bare cosmological constant exactly cancel against the 
background expansion of the quadratic term. The theory undergoes 
a mathematical reduction, becoming isomorphic to pure 
quadratic $R^2$ gravity, but rigidly shifted to a non-zero background curvature 
$R_0$.

Thus, the emergence of the arbitrary $c_2/r^2$ 
integration constant is a direct, shifted analogue of the 
degeneracy in pure $R^2$ gravity \cite{Kehagias2015}, where exact 
solutions with arbitrary $1/r^2$ terms are permitted 
strictly when the scalar curvature vanishes ($R=0$). By fine-tuning 
the bare cosmological constant, this pure quadratic decoupling 
is systematically reconstructed at a finite de Sitter-like curvature value. 

\subsection{Degenerate Wald entropy}

While the formal geometry mimics a charged black hole, 
the thermodynamic properties of the degenerate spacetime 
strictly reflect this complete loss of dynamics. 
In any diffeomorphism-invariant 
generalized theory of gravity, the entropy of a black hole is 
not merely proportional to its horizon area, but is rather 
governed by the Wald entropy formula \cite{Wald1993, Vivek1994}, 
which accounts for the higher-order curvature terms. 

For $f(R)$ gravity theories \cite{Brevik2004,Brustein2009}, 
the Wald entropy evaluated at the event horizon $r_H$ is given by
\begin{equation}
S_{\text{Wald}} = \frac{\mathcal{A}_H}{4G} f'(R_H),
\end{equation}
where $\mathcal{A}_H = 4\pi r_H^2$ is the horizon area, and 
$G$ is the bare Newton's constant. For the standard, dynamical 
Schwarzschild-de Sitter branch in our 
$\Lambda$-vacuum model, $1 - 8\alpha \Lambda > 0$ holds, 
and the entropy is finite and strictly positive. 
However, for the degenerate branch with $f'(R) = 0$ globally, 
the entropy identically vanishes at the event horizon. 
Therefore, the black holes in this fine-tuned family possess 
exactly zero Wald entropy,
\begin{equation}
S_{\text{Wald}} = 0.
\end{equation}

Recall that the Wald entropy is formally derived as 
a Noether charge evaluated on the bifurcation surface 
of a Killing horizon. Therefore, rather than a paradox, the result $S_{\text{Wald}} = 0$ 
is most accurately interpreted as the macroscopic 
vanishing of this Noether charge at the $f'(R)=0$ limit, 
acting as a vital thermodynamic consistency check for the breakdown of the 
standard dynamical degrees of freedom. 
While one might be tempted to appeal to the standard 
Boltzmann relation 
($S = k_B \ln \Omega$) to suggest this zero entropy indicates a spacetime 
macrostate possessing only a single valid microstate ($\Omega=1$), 
such an extrapolation is highly speculative and mathematically unjustified in this context. 
As demonstrated by the thermodynamics of extremal black holes and 
topological gravity models, a vanishing macroscopic entropy does not 
necessarily imply a trivial quantum state counting. The zero entropy found 
here is strictly a classical and semi-classical thermodynamic pathology 
indicative of the degenerate strong-coupling 
boundary.\footnote{That such degeneracy 
may eventually be removed by quantum corrections was demonstrated via the 
one-loop analysis for the de Sitter case in Ref.\ \cite{Cognola2005}.}

It is worth noting that the phenomenon of black hole 
horizons possessing a finite temperature but an identically 
vanishing entropy is not entirely unprecedented in 
higher-order curvature theories. Similar thermodynamic 
behavior has been rigorously observed in other exact solutions within 
modified gravity. For instance, Lifshitz black holes derived 
in four-dimensional pure $R^2$ gravity \cite{Cai2009Lifshitz}, 
as well as specific classes of black holes in Lovelock gravity 
\cite{Cai2010Lovelock}, similarly exhibit non-vanishing 
temperatures alongside exactly zero entropy and mass. 
In those contexts, much like the exact $f'(A)=0$ degeneracy 
explored here, the strict vanishing of the effective gravitational 
degrees of freedom at the horizon perfectly extinguishes 
the horizon entropy, further highlighting the deep connection 
between mathematically degenerate boundaries and pathological 
black hole thermodynamics.

\subsection{Strong coupling and breakdown 
of the test-particle approximation}

The underlying physical reason for both the unconstrained $c_2$ parameter 
and the identically zero entropy lies in the effective gravitational 
coupling of the theory. In $f(R)$ gravity, the effective, dynamic 
Newton's constant is defined as
\begin{equation}
G_{\text{eff}} = \frac{G}{f'(R)}.
\end{equation}
The degenerate solution lives precisely on the singular boundary where 
$f'(R) = 0$, implying that the effective coupling strictly diverges 
($|G_{\text{eff}}| \to \infty$). 

In the dynamically equivalent scalar-tensor representation 
of Starobinsky gravity, the theory can be mapped to Einstein 
gravity coupled to a scalar field (the ``scalaron''). At the degenerate 
point $f'(R) = 0$, the scalaron field ceases to propagate 
normally, and its interaction vertices diverge. 

Physically, this means that while the exact static background 
$h(r) = 1 - 2M/r + c_2/r^2 - r^2/(24\alpha)$ is a valid mathematical 
solution to the trace-decoupled field equations, the standard test-particle 
approximation completely fails. Any physical probe with non-zero mass 
introduced into this spacetime will conceptually induce an infinite backreaction. 
Thus, linear perturbation theory completely breaks down, and the spacetime 
cannot be physically realized or dynamically evolved. The appearance of 
the arbitrary constant $c_2$ is not a sign of a robust new geometric state, 
but rather a direct symptom of the field equations losing their predictive 
power at this extreme strong-coupling limit.

\subsection{Perturbative behavior near the degenerate boundary}
\label{sec:perturbation}

To elaborate on the absolute physical fragility 
of the $c_2/r^2$ branch, it is instructive to 
examine the spacetime strictly in the neighborhood of the 
fine-tuned boundary, rather than exactly on it. We parameterize a small 
departure from the exact degenerate state by introducing a dimensionless 
perturbation parameter, $\epsilon$, such that the bare cosmological 
constant is given by
\begin{equation}
\Lambda = \frac{1}{8\alpha} (1 + \epsilon),
\end{equation}
where $0 < |\epsilon| \ll 1$. Assuming the spacetime maintains 
a constant scalar curvature $R = A$, the vacuum trace constraint 
given by Eq.\ (\ref{eq:scalarConstraint}) once again  
necessitates $A = -4\Lambda$. Under this perturbation, the 
global scalar curvature becomes
\begin{equation}
A = -\frac{1}{2\alpha} (1 + \epsilon).
\end{equation}

We now insert this perturbed parameter into our factorized 
master equation, Eq.\ (\ref{eq:4dStaroMODconstantR}). 
The prefactor evaluates exactly to $1 - 8\alpha\Lambda = -\epsilon$. 
Consequently, the tensor field equations for the perturbed spacetime 
take the form,
\begin{equation}
-\epsilon \left( R_{\mu\nu} + \Lambda  g_{\mu\nu} \right) = 0.
\end{equation}
Because we are explicitly evaluating the perturbative regime 
where $\epsilon \neq 0$, we are mathematically permitted to 
divide the entire equation by the perturbation parameter. 
By doing so, the decoupling of the field equations 
instantaneously fails, and the system rigidly snaps back to 
the standard general relativistic vacuum field equations, 
$R_{\mu\nu} + \Lambda  g_{\mu\nu} = 0$. Our general 
metric Ansatz from Eq.\ (\ref{eq:h(r)}) identically forces 
the purely geometric arbitrary term to vanish, 
yielding the strict requirement,
\begin{equation}
c_2 = 0.
\end{equation}
This discontinuous collapse sharply restricts 
the physical phase space of the theory. 
It demonstrates that the $c_2 \neq 0$ 
spacetime cannot serve as a continuous background 
geometry, $g_{\mu\nu}^{(0)}$, for physical perturbations. 
In standard perturbation theory (e.g., \cite{Radhakrishnan2026}), 
one expects a perturbed solution to smoothly approach 
the background as the perturbation parameter vanishes 
($\epsilon \to 0$). However, the moment the bare 
cosmological constant deviates from $\Lambda = 1/(8\alpha)$ 
by even an infinitesimal amount, the $c_2/r^2$ term is entirely 
forbidden. The exact solution does not smoothly fade into the 
surrounding parameter space; rather, it discontinuously collapses 
into the standard Schwarzschild-de Sitter family. 

This simple perturbative analysis elegantly pinpoints the physical 
pathologies inherent to the degenerate boundary. In this 
perturbed regime, the effective gravitational coupling becomes
\begin{equation}
G_{\text{eff}} = -\frac{G}{\epsilon}.
\end{equation}
As the system is tuned toward the exact boundary 
($\epsilon \to 0$), $|G_{\text{eff}}| \to \infty$, confirming the infinite 
strong-coupling limit. Moreover, positively crossing the 
boundary ($\epsilon > 0$) forces the effective coupling to 
become strictly negative. This triggers a ghost 
instability where gravity becomes fundamentally repulsive 
and the graviton acquires a negative kinetic term. Thus, the 
degenerate solution is not merely an isolated point in parameter 
space; it is a physically inaccessible mathematical 
wall separating standard gravitation from an intensely 
pathological regime.

\subsection{Probe motion and the breakdown 
of linear perturbation theory}

As a direct consequence of this infinite strong-coupling limit, 
the physical viability of treating local matter distributions 
as standard test particles is invalidated. 
Given the exact analytical form of the metric $h(r)$ in the 
degenerate branch, one could formally compute the Christoffel 
symbols and define mathematical geodesic trajectories. 
However, the geodesic equation implicitly relies on the 
test-particle approximation, assuming that the mass $m$ 
of the probe is sufficiently small that its linear backreaction 
on the background geometry, 
$\delta g_{\mu\nu} \propto G_{\text{eff}} m$, is negligible.

Because the effective gravitational coupling diverges 
across the entirety of the degenerate spacetime, 
the linear perturbation expansion completely fails. Any 
physical probe with a finite, non-zero mass yields a formally 
divergent linear backreaction. Consequently, the standard 
test-particle approximation breaks down entirely. 
This implies that one cannot operationally measure the mass 
parameter $M$ of this geometry via standard Keplerian orbits 
of local probes. Instead, $M$ must be understood purely as 
a formal geometric integration constant characterizing the intrinsic 
curvature of the mathematical vacuum, rather than its measurable extrinsic 
effect on matter.

It is important to qualify that this perturbative divergence demonstrates 
the failure of the classical linear expansion, rather than 
serving as a rigorous mathematical proof of absolute non-linear instability. 
In strongly coupled regimes, while linear perturbation theory fails, 
exact non-perturbative solutions incorporating matter might 
theoretically exist. To definitively prove an absolute instability 
of the spacetime, one would need to linearize the full field 
equations around this background and demonstrate the absence 
of a well-posed Cauchy problem, an analysis that falls outside 
the scope of the present derivation. 

Therefore, the exact solution derived here is most accurately 
classified as the identification of a singular parameter locus, 
or a strong-coupling boundary, intrinsic to the Starobinsky model. 
It pedagogically demonstrates that exactly at the degenerate point $f'(R)=0$, 
the vacuum theory cannot dynamically support standard matter 
within the linearized regime, beautifully illustrating why $f(R)$ theories require 
healthy $f'(R)>0$ limits to function.

\subsection{Global structure, extremal limits, and the $c_2 < 0$ regime}

To characterize the formal geometry of the solution, 
one can examine the global horizon structure defined by 
the roots of the metric function, $h(r) = 0$. 
The roots of our degenerate branch are determined by the 
quartic polynomial,
\BEq
r^4 - 24\alpha r^2 + 48\alpha M r - 24\alpha c_2 = 0.
\EEq 

When the geometric integration constant is strictly positive ($c_2 > 0$), 
the mathematical structure of the spacetime is isomorphic 
to the standard Reissner-Nordstr\"om-de Sitter (RN-dS) geometry, 
where we make the formal identification $Q^2 \leftrightarrow c_2$ 
and $\Lambda_{\text{eff}} \leftrightarrow 1/(8\alpha)$. Depending 
on the relative magnitudes of $M$ and $c_2$, this formal regime admits 
up to three horizons (via Vieta’s formulas, due to the absence of
the cubic term): an inner Cauchy horizon ($r_-$), an outer 
event horizon ($r_+$), and a cosmological horizon ($r_c$). Extremal 
limits occur identically to the RN-dS case when the roots degenerate 
($h(r)=0$ and $h'(r)=0$); for example, the extremal black hole limit 
$r_- = r_+$, or the Nariai limit $r_+ = r_c$.

However, because $c_2$ is a geometric integration 
constant arising from the decoupled $f'(R)=0$ boundary, it is not 
bound by the positivity constraint of a physical Maxwell field 
($Q^2 \ge 0$). The regime where $c_2 < 0$ introduces causal 
structures with no standard electro-vacuum counterpart. 
When $c_2 < 0$, the $c_2/r^2$ term dominates as $r \to 0$, 
driving $h(r) \to -\infty$. Consequently, the inner Cauchy horizon 
($r_-$) is eliminated. If the formal mass parameter $M$ 
is sufficiently large, the mathematical spacetime possesses 
an event horizon and a cosmological horizon, 
but the central singularity at $r=0$ 
becomes spacelike (analogous to the standard Schwarzschild 
singularity) rather than timelike. Conversely, if $M$ is sufficiently 
small or vanishes, $h(r)$ remains strictly negative for all $r < r_c$, 
exposing $r=0$ as a formal naked singularity.

While one could construct maximally extended Penrose 
conformal diagrams for these various sub-cases, they would be 
topologically identical to the well-cataloged diagrams for RN-dS 
(for $c_2 > 0$) or a modified Schwarzschild-de Sitter geometry 
(for $c_2 < 0$). Physically, however, such constructions are purely 
formal mathematical exercises. As established in the preceding 
sections, the divergent effective coupling at the $f'(R)=0$ boundary 
triggers a breakdown of linear perturbation theory. Because any 
physical probe or classical perturbation would induce a divergent 
backreaction, the maximally extended global causal structure (which 
implicitly assumes the background geometry remains stable across 
infinite affine parameterization) is physically unrealizable. 
This emphasizes that the degenerate spacetime serves as a 
mathematical boundary state rather than a globally traversable 
physical background.

\subsection{Cosmological viability and matter instability}

These physical pathologies, ranging from the failure of 
linear perturbation theory to the divergent backreaction of local 
probes, can be seamlessly understood by framing the exact 
solution within the established theoretical viability conditions for 
generalized $f(R)$ gravity. For any higher-derivative gravitational 
theory to represent a physically viable extension of general relativity, 
it must satisfy a strict set of stability criteria governing both the vacuum 
and its coupling to matter. For example, as was shown long ago in 
\cite{Nojiri2003}, canceling matter instability in pure $R^2$ gravity 
requires the inclusion of a $1/R$ term. Subsequently, Faraoni 
\cite{Faraoni2006}, in his analysis of the Dolgov-Kawasaki instability 
\cite{Dolgov2003}, showed that a viable $f(R)$ model must possess 
a positive second derivative, $f''(R) > 0$ (cf.\ \cite{Amendola2007}). 
This condition ensures that the effective mass squared of the 
scalaron remains real and positive. If $f''(R) < 0$, the theory suffers from 
a matter instability where the scalaron field becomes tachyonic, 
leading to the rapid destabilization of any macroscopic matter distribution 
introduced into the spacetime. In addition, as detailed in studies 
of viable cosmological structure formation by Pogosian and Silvestri 
\cite{Pogosian2008}, the theory must also satisfy a ghost-free condition 
requiring a positive first derivative, $f'(R) > 0$. This ensures that the 
effective gravitational coupling ($G_{\text{eff}} = G/f'(R)$) remains 
finite and positive.

For the pure Starobinsky vacuum explored here, governed by 
$f(R) = R + \alpha R^2 + 2\Lambda$, the stability criteria remain 
applicable, as the addition of a bare cosmological constant vanishes 
upon differentiation. The second derivative evaluates to a constant, 
$f''(R) = 2\alpha$. Therefore, satisfying Faraoni's matter stability 
criterion requires the coupling parameter to be positive ($\alpha > 0$). 
In the context of our degenerate branch, this enforces a positive bare 
cosmological constant, $\Lambda = 1/(8\alpha) > 0$, ensuring the 
formal spacetime is de Sitter-like. 

However, while one can tune the theory to avoid the Dolgov-Kawasaki 
instability via $\alpha > 0$, the exact formal geometry of the degenerate 
branch corresponds to $f'(R) = 0$ globally. By sitting precisely 
on the boundary that violently violates the ghost-free 
viability condition of Pogosian and Silvestri, the spacetime is rendered 
physically unviable as a macroscopic phase. Even if the underlying scalaron 
is formally non-tachyonic ($f''(R) > 0$), this exact violation of the ghost-free 
boundary is what triggers the infinite strong coupling and the failure of 
linear perturbation theory discussed in the preceding subsections. 
Thus, while the $c_2/r^2$ geometry is an exact solution of the 
trace-decoupled field equations, it represents a degenerate state that 
cannot support viably evolving matter or stable spacetime perturbations.

The necessity of enforcing these physical viability conditions 
(particularly the ghost-free requirement $f'(R) > 0$ and the matter stability 
criterion $f''(R) > 0$) extends beyond the classical stability of exact 
black hole solutions. For instance, in the context of the AdS/CFT 
correspondence, quadratic curvature corrections influence the holographic 
transport properties of strongly coupled dual field theories. Recent work on 
four-dimensional AdS black branes supplemented by $R^2$ modifications 
demonstrates that thermodynamic stability and causality bounds on the 
quadratic coupling parameter are linked to the shear viscosity-to-entropy 
density ratio ($\eta/s$), potentially driving it below the Kovtun-Starinets-Son 
(KSS) bound \cite{Golmoradifard2025AdSViscosity}. Thus, analyzing exact 
pathological boundaries and the breakdown of stability conditions in $f(R)$ 
models provides complementary insights, ensuring that constraints on curvature 
couplings remain theoretically consistent across classical gravity and 
holographic applications.

\subsection{Conjectural resolution of the degeneracy via quantum corrections}
\label{sec:quantum}

As demonstrated above, the exact fine-tuning $\Lambda = 1/(8\alpha)$ 
drives the effective gravitational coupling to infinity ($|G_{\text{eff}}| \to \infty$). 
In such a strongly coupled regime, purely classical considerations become 
incomplete, and quantum corrections to the spacetime geometry can no 
longer be neglected. Drawing on established literature, we can anticipate 
the semi-classical behavior of the system. The inclusion of one-loop quantum 
effects modifies the classical action, yielding an effective Lagrangian 
$f_{\text{eff}}(R) = f(R) + \Gamma^{(1)}(R)$, where $\Gamma^{(1)}(R)$ 
encapsulates the one-loop effective action. The generic structure of such 
quantum corrections in $f(R)$ gravity has been evaluated in 
constant-curvature de Sitter backgrounds by Cognola \textit{et al.}\ 
\cite{Cognola2005}, demonstrating that quantum fluctuations introduce 
non-trivial logarithmic and higher-order curvature terms to the effective action.

While the classical derivative identically vanishes at the fine-tuned 
boundary ($f'(R_0) = 0$), there is no known fundamental symmetry protecting this 
classical degeneracy against quantum contributions. Generically, one 
expects the derivative of the one-loop correction will not identically 
vanish at this specific scalar curvature ($\Gamma^{(1)\prime}(R_0) \neq 0$). 
If this expectation holds, the fully quantized effective derivative becomes non-zero,
\begin{equation}
f_{\text{eff}}'(R_0) 
= 
f'(R_0) + \Gamma^{(1)\prime}(R_0) 
\approx 
\Gamma^{(1)\prime}(R_0) \neq 0.
\end{equation}
By shifting the effective derivative away from zero, such quantum 
corrections would systematically lift the classical $R_0$-degeneracy. Just as an 
infinitesimal classical shift in the bare cosmological constant rigidly forbids 
the $c_2/r^2$ arbitrary term, the generic addition of quantum corrections 
provides a dynamical mechanism to break the exact decoupling of the 
field equations. Therefore, we hypothesize that in a realistic semi-classical 
theory, the fully decoupled algebraic constraint is radiatively lifted. 
This strongly suggests that quantum fluctuations dynamically resolve the pathology 
of the degenerate boundary, inherently preventing the formation of the geometric 
mimicker and ensuring the spacetime collapses back into the standard, 
non-degenerate Schwarzschild-de Sitter geometry.

\section{Conclusion}

In this paper, we have explored the exact boundary limits of the 
strong-field regime in the four-dimensional Starobinsky $R+\alpha R^2$ model, 
specifically focusing on exact, static, spherically-symmetric vacuum spacetimes. 
By carefully analyzing the trace of the modified field equations, we mapped 
a strict mathematical degeneracy accessed via the inclusion of a bare 
cosmological constant. We demonstrated that fine-tuning 
this cosmological constant to exactly $\Lambda = 1/(8\alpha)$ 
rigidly forces the background scalar curvature to a constant value 
of $R = -1/(2\alpha)$. At this precise curvature, the derivative 
of the gravitational Lagrangian with respect to the Ricci scalar, 
$f'(R) = 1+2\alpha R$, identically vanishes across the entire 
spacetime.

This global vanishing of $f'(R)$, which occurs at a finite, 
non-zero scalar curvature $R=-1/(2\alpha)$, effectively 
decouples the higher-derivative constraints on the traceless 
Ricci tensor. Unlike pure quadratic gravity models where 
decoupling requires a strictly Ricci-flat geometry ($R=0$), 
this constant-curvature background permits the exact 
integration of a mathematically degenerate, two-parameter family of 
formal spacetimes. Distinct from the standard, dynamical Schwarzschild-de Sitter 
geometry, this analytical solution features a fundamentally unconstrained $1/r^2$ 
integration artifact. Macroscopically, this purely geometric term 
allows the mathematically degenerate Starobinsky vacuum to perfectly mimic 
the formal metric of a Reissner-Nordstr\"om black hole. This provides 
an exact pedagogical demonstration of how higher-order curvature terms, 
when pushed to a decoupled singular limit, can formally replicate 
the metric contribution of a physical $U(1)$ gauge field.

However, the physical and thermodynamic analysis of this 
exact mathematical solution reveals the severe pathologies 
inherent to the $f'(R)=0$ boundary. Drawing on the Wald 
entropy formalism, we showed that because the effective 
gravitational coupling strictly diverges ($|G_{\text{eff}}| \to \infty$), 
any event horizon in this fine-tuned spacetime evaluates to 
identically zero thermodynamic entropy; a result that acts as a vital 
consistency check reflecting the macroscopic loss of dynamical degrees of freedom. 
Furthermore, this infinite strong-coupling limit triggers a complete breakdown 
of the standard test-particle approximation. The divergent effective coupling 
indicates that any physical mass introduced into the spacetime 
will formally induce an infinite linear backreaction, immediately destabilizing 
the background. Within the dynamically equivalent scalar-tensor formulation, 
this geometrically manifests as a strict strong-coupling limit of the scalaron field.

Ultimately, the degenerate solution presented here serves as an 
instructive mathematical boundary marker and a vital pedagogical tool 
for illustrating why the strict physical viability conditions of modified gravity 
(such as the ghost-free $f'(R)>0$ requirement) must be rigidly enforced. 
It highlights that the highly non-linear nature of $f(R)$ theories 
can harbor exact analytical spacetimes that extend beyond standard 
general relativity, but it powerfully demonstrates that algebraic 
exactness does not guarantee physical reality or predictive power. Spacetimes 
residing at the $f'(R)=0$ boundaries of the theory's phase 
space are physically unviable within classical linear perturbation theory and are 
expected to be fundamentally unstable against radiative quantum corrections. 
Nevertheless, while standard classical and perturbative techniques fail at this 
degenerate locus, it remains an open theoretical question whether fully non-perturbative 
treatments, exact UV completions, or extended gravitational actions might one day 
regularize this strong-coupling limit. Future theoretical work might explore 
whether such advanced frameworks can radiatively resolve the zero-entropy 
degeneracy, or definitively confirm that these exact, purely geometric states 
represent universally inaccessible, pathological endpoints in the phase space 
of higher-order gravity.

\begin{acknowledgments}

The author thanks Sergei Odintsov for helpful correspondence.

\end{acknowledgments}

\end{document}